\documentclass[aps,prl,twocolumn,superscriptaddress,showpacs,amsmath,amssymb]{revtex4}
\usepackage[usenames,dvipsnames]{color}
\usepackage{epsf}
\usepackage{bm}
\usepackage{dcolumn}
\usepackage{latexsym}
\usepackage{amsmath}
\usepackage{amsfonts}
\usepackage{amssymb}
\usepackage{graphicx}
\usepackage[active]{srcltx}
\newcommand{\be}{\begin{eqnarray}}
\newcommand{\ee}{\end{eqnarray}}

\newcommand{\bi}{\hat{b}}

\def\rmd{{\rm{d}}}
\def\rmi{{\rm{i}}}
\def\rme{{\rm{e}}}
\def\bm#1{\mbox{\boldmath{$#1$}}}
\def\bi#1{{\bm #1}}

\definecolor{darkred}{rgb}{.8,0,0}

\definecolor{darkblue}{rgb}{0,0,.7}

\begin{document}
%
 \title{
 Special Relativity induced by Granular Space}
%
%
\author{Petr Jizba}
\email{p.jizba@fjfi.cvut.cz}
\affiliation{FNSPE, Czech Technical
University in Prague, B\v{r}ehov\'{a} 7, 115 19 Praha 1, Czech Republic\\}
\affiliation{ITP, Freie Universit\"{a}t Berlin, Arnimallee 14
D-14195 Berlin, Germany}
\author{Fabio Scardigli}
\email{fabio@phys.ntu.edu.tw}
\affiliation{Dipartimento di Matematica, Politecnico di Milano, Piazza Leonardo da Vinci 32, 20133 Milano, Italy\\}
\affiliation{Yukawa Institute for Theoretical Physics, Kyoto
University, Kyoto 606-8502, Japan}
\begin{abstract}
\vspace{1mm}
We show that the special relativistic dynamics when combined with quantum mechanics and the concept of superstatistics can be interpreted as arising from two interlocked non-relativistic stochastic processes that operate at different energy scales. This interpretation leads to Feynman amplitudes that are in the Euclidean regime identical to transition probability
of a Brownian particle propagating through a granular space. Some kind of spacetime granularity could be held responsible for the emergence at larger scales of various symmetries.
For illustration we consider also the dynamics and the propagator of a spinless relativistic particle. Implications for doubly special relativity, quantum field theory, quantum gravity and cosmology are discussed.
\end{abstract}
%
\maketitle
%
%

{\em {Introduction}.}~---~The concept of ``emergence" plays an important role in quantum field theory
and, in particular, in particle and condensed matter physics, since it embodies the essential feature of systems with several interlocked time scales. In these systems, the observed macroscopic-scale dynamics and related degrees of freedom differ
drastically from the actual underlying microscopic-scale physics~\cite{Laughlin:05}.
Superstatistics provides a specific realization of this paradigm: It predicts that the emergent behavior can
be often regarded as a {\em superposition} of several statistical systems that operate at different spatio-temporal
scales~\cite{Beck:01,Gell-Mann:11}. In particular, many applications have recently been  reported,
in hydrodynamic turbulence~\cite{Reynolds}, turbulence in quantum liquids~\cite{Beck:12},
pattern forming systems~\cite{Daniels:04} or scattering processes in
high-energy physics~\cite{Wilk:08}.

The essential assumption of the superstatistics scenario is the existence
of sufficient spatio-temporal scale separations between relevant
dynamics within the studied system so that the system has
enough time to relax to a local equilibrium state and to stay within
it for some time. In practical applications one is typically concerned with two scales.
Following~Ref.~\cite{Beck:01}, we consider an intensive parameter
$\zeta$ that fluctuates on a much larger time scale than
the typical relaxation time of the local dynamics. The random
variable $\zeta$ can be in practice identified, e.g., with the inverse temperature~\cite{Beck:01,Gell-Mann:11},
friction constant~\cite{beckIII}, volatility~\cite{JK1} or einbein~\cite{JK2}.
On intuitive ground, one may understand the superstatistics by using the adiabatic Ansatz. Namely,
the system under consideration, during its evolution, travels within its state space $X$ (described by state variable $A\in X$)
which is partitioned into small cells characterized by a sharp value of $\zeta$. Within each cell, the system is described
by the conditional distribution $p(A|\zeta)$. As $\zeta$ varies adiabatically
from cell to cell, the joint distribution of finding the system with a sharp value of $\zeta$ in the state $A$
is $p(A,\zeta) = p(A|\zeta)p(\zeta)$ (Bayes theorem).
The resulting macro-scale (emergent) statistics $p(A)$ for finding system in the state $A$ is obtain by eliminating the nuisance parameter $\zeta$ through marginalization, that is
\begin{eqnarray}
p(A) \ = \ \int p(A|\zeta)p(\zeta)\ \! d\zeta \, .
\label{I.1.a}
\end{eqnarray}
Interestingly enough, the sufficient time scale
separation between two relevant dynamics in a studied
system allows to qualify superstatistics as a form of slow
modulation~\cite{Allegrini:06}.

In this Letter, we recast the Feynman transition amplitude of a relativistic scalar particle into a form, which (after being analytically continued to imaginary times) coincides with a superstatistics marginal probability (\ref{I.1.a}).
The derivation is based on the L\'{e}vy--Khinchine theorem for infinitely divisible distributions~\cite{feller66,Veillette:11}, and
for illustration we consider the dynamics and the propagator of a Klein--Gordon (i.e., neutral spin$-0$) particle. Our reasonings can be also extended to charged spin$-0$,
spin$-\frac{1}{2}$, Proca's spin$-1$ particles and to higher-spin particles phrased via the Bargmann--Wigner wave equation~\cite{JK2}. Further generalization to external electromagnetic potential
has been reported in Refs.~\cite{JK2,JS}.

We also argue that the above formulation can be looked at as
if the particle would
randomly propagate (in the sense of Brownian motion) through an inhomogeneous or granular medium (``vacuum")~\cite{JS}.

Our argument is based upon a recent
observation~\cite{JK1,JK2,JS} that the Euclidean path integral (PI) for
relativistic particles may be interpreted as describing a doubly-stochastic process that operates at two separate
spatio-temporal scales.
The short spatial scale, which is much smaller than
particle's Compton length $\lambda_{_C} = 1/mc$ ($\hbar = 1$), describes a Wiener (i.e., Galilean relativity)
process with a sharp (Galilean-invariant) Newtonian mass.
The large spatial scale, which is of order $\lambda_{_C}$, corresponds to distances over which the fluctuating Newtonian mass changes appreciably. At scales much larger than $\lambda_C$ the particle evolves according to a genuine
relativistic dynamics, with a sharp value of the mass coinciding with the Einstein rest mass.
Particularly striking is the fact that when we average the particle's velocity over the
structural correlation distance (i.e., over particle's $\lambda_{_C}$) we obtain the velocity of light $c$.
So the picture that emerges from this analysis is that the particle (with a non-zero mass!)
propagates over the correlation distance $~ \lambda_{_C}$
with an average velocity $c$, while at larger
distance scales (i.e., when a more coarse grained view is taken) the particle propagates as a relativistic particle with a sharp mass and an average velocity that is subluminal.
Quite remarkably, one can observe an identical behavior in
the well-known Feynman's checkerboard PI~\cite{FH, Schulman-Jacobson} to which the transition amplitude (\ref{I.1.a}) reduces in the case of a relativistic Dirac fermion in $1+1$ dimensions~\cite{JK2,JS}.

A considerably expanded presentation including the issue of reparametrization
invariance, bibliography, and proofs of the main statements and formulas is
given in a companion paper~\cite{JS}.


{\em {Superstatistics path integrals.}}~---~When a conditional probability density
function (PDF) is formulated through a PI, then it satisfies
the Einstein-–Smoluchowski equation (ESE) for continuous Markovian
processes, namely~\cite{PI}
\begin{eqnarray}
 p(y, t''|x,t)=  \int_{-\infty}^{\infty} \rmd z\ \!
 p(y,
t''|z, t') p(z, t' |x,t)\, ,
\label{II1a}
\end{eqnarray}
with $t'$ being any time between $t''$ and $t$.
Conversely, any transition probability satisfying ESE possesses a PI
representation~\cite{FH}.
In physics one often encounters probabilities formulated as a
superposition of PI's,
\begin{eqnarray} \label{2aa}
&&\mbox{\hspace{-5mm}}\wp(x',t'|x,t) \nonumber \\
&&\mbox{\hspace{-5mm}}= \! \int_{0}^{\infty} \rmd \zeta \  \omega(\zeta,T)
\int_{x(t) = x}^{x(t') = x'}[\rmd x \ \!\rmd p]\
e^{\int_{t}^{t'} \rmd \tau \left(\footnotesize{\rmi} p\dot{x} - \zeta
H(p,x) \right)}\, . \nonumber \\
\end{eqnarray}
Here $\omega(\zeta,T)$ with $T=t'-t$ is a normalized PDF
defined on ${\mathbb{{R}}}^+\!\!\times {\mathbb{{R}}}^+$.
The form (\ref{2aa}) typically appears in non-perturbative approximations to
statistical partition functions, in polymer physics, in financial markets,
in systems with reparametrization invariance, etc.  The random variable
$\zeta$ is then related to the inverse temperature,
coupling constant, volatility, vielbein, etc.

The existence of different time scales
and the flow of the information from slow to fast degrees
of freedom typically creates the irreversibility on the macroscopical
level of the description. The corresponding
information thus is not lost, but passes in a form
incompatible with the Markovian description. The most general class of distributions $\omega(\zeta,T)$ on  ${\mathbb{{R}}}^+\!\!\times {\mathbb{{R}}}^+$
for which the superposition of Markovian processes remain Markovian, i.e., when also $\wp({x}',t'|{ x},t)$
satisfies the ESE (\ref{II1a}), was found in Ref.~\cite{JK1}. The key is to note that in order to have (\ref{II1a}) satisfied by $\wp$, the rescaled PDF $w(\zeta,T)\ \equiv\  \omega (\zeta/T,T)/T\,$ should satisfy the ESE for homogeneous Markovian process
\begin{eqnarray}
w(\zeta,t_1+ t_2)  \ = \ \int_0^\zeta \rmd \zeta' \ \! w(\zeta',t_1) w(\zeta-\zeta',t_2)\, .
\end{eqnarray}
Consequently the Laplace image fulfills the functional equation with $t_1,t_2 \in \mathbb{R}^+$. By assuming continuity in  $T$, it follows that the multiplicative semigroup $\tilde{w}(p_{_\zeta},T)_{T\geq 0}$  satisfies $\tilde{w}(p_{_\zeta},T) = \{\tilde{w}(p_{_\zeta},1)\}^T$. From this we see that the distribution of $\zeta$ at $T$ is completely determined by the distribution of $\zeta$ at $T=1$. In addition,
because  $\tilde{w}(p_{_\zeta},1) = \{\tilde{w}(p_{_\zeta},1/n)\}^n$ for any $n \in \mathbb{N}^+$, $w(\zeta,1)$ is infinitely divisible. The L\'{e}vy--Khinchine theorem~\cite{feller66,Veillette:11} then ensures that $\log \tilde{w}(p_{_\zeta},T) \equiv -T F(p_{_\zeta})$
must have the generic form
\begin{eqnarray}
\log\tilde{w}(p_{_\zeta},T) \ = \  -T\left(\alpha p_{_\zeta} + \int_{0}^{\infty}(1 - e^{-p_{_{\zeta}} u}) \nu(\rmd u)\right)\! ,
\label{II.5.a}
\end{eqnarray}
where $\alpha\geq 0$ is a drift constant and $\nu$ is some non-negative measure on $(0,\infty)$ satisfying $\int_{\mathbb{R}^+} \mbox{min}(1,u)\nu(\rmd u) < \infty$.
Finally the Laplace inverse of
$\tilde w(p_{_\zeta},T)$ yields $\omega(\zeta,T)$.
Once $\omega(\zeta,T)$ is found, then  $\wp({x}',t'|{x},t)$
possesses a PI representation on its own.
What is the form of the new Hamiltonian? To this end we rewrite (\ref{2aa})
in Dirac operator form as~\cite{JK1}
\begin{eqnarray}
&&\mbox{\hspace{-5mm}}\wp(x',t'|x,t) \ = \ \langle x' | \int_{0}^{\infty} \!\!\!\rmd \zeta \ \! w(\zeta, T) e^{-\zeta \hat{H}}|x\rangle \nonumber \\
&&\mbox{\hspace{-5mm}}= \ \langle x' |\{ \tilde{w}(\hat{H},1) \}^{T} |x \rangle
\ = \  \langle x' |e^{-TF(\hat{H})} |x \rangle\, .
\label{6aa}
\end{eqnarray}
Hence, the identification $\mathcal{H}({\bi p},{\bi x}) \ = \ F(H({\bi p},{\bi
x}))$ can be made. Here one might worry about the
operator-ordering problem.
For our purpose it suffices to note that when $H$ is ${\bi x}$-independent, the
former relation is exact.  In more general instances the Weyl ordering
is a natural choice because in this case the required mid-point rule follows automatically and one does not need to invoke the gauge invariance~\cite{JK1,fiorenzo}.
In situations when other non-trivial configuration space symmetries (such as non-holonomic symmetry) are required, other orderings might be more physical~\cite{JK1}.
%


{\em {Emergent Special Relativity}.}~---~
The Feynman transition amplitudes (or better its Euclidean version --- transition probabilities)
naturally fits into the structure of superstatistics PI's discussed above.

Note first that the choice $\alpha=0$
and $\nu(\rmd u) = 1/(2\sqrt{\pi}u^{3/2}) \rmd u$ leads to
$F(p_{_\zeta}) = \sqrt{p_{_\zeta}}$. This identifies $w(\zeta, T)$ with the (unshifted) L\'{e}vy distribution with the scale parameter $T^2/2$. Moreover, when
$H({\bi p},{\bi x}) = {\bi p}^2c^2 + m^2c^4$  then (\ref{2aa}) can be cast into the form (see also Refs.~\cite{JK1,JK2,JS})
\begin{widetext}
\begin{eqnarray}
 &&\mbox{\hspace{-15mm}}\int_{\footnotesize{\bi x}(0)\ \! =
 \ \! \footnotesize{\bi x}}^{\footnotesize{\bi
x}(T) \ \! = \ \! \footnotesize{\bi x}'} \ \![\rmd{\bi x} \ \!\rmd
{\bi p}] \ \! \exp\left\{\int_{0}^{T} \!\!\rmd \tau \
\!\left[\rmi {\bi p}\cdot \dot{\bi x} \ - \
c\sqrt{{\bi p}^2 + m^2 c^2}\right]\right\} \nonumber \\[2mm]
&&\mbox{\hspace{15mm}}= \  \int_{0}^{\infty}\!\!\rmd \mathfrak{m} \
\! \textmd{f}_{\frac{1}{2}}\!\left(\mathfrak{m}, Tc^2, Tc^2m^2\right)
\int_{\footnotesize{\bi x}(0)\ \! = \ \! \footnotesize{\bi x}}^{\footnotesize{\bi
x}(T) \ \! = \ \! \footnotesize{\bi x}'} \ \![\rmd{\bi x} \ \!\rmd
{\bi p}] \ \! \exp\left\{\int_{0}^{T} \!\!\rmd \tau\
\!\left[\rmi {\bi p}\cdot \dot{\bi x} \ - \  \frac{{\bi p}^2}{2
\mathfrak{m}} \ - \  m c^2 \right]\right\}\, ,
\label{22a}
\end{eqnarray}
\end{widetext}
where $t'-t  = T$, and
\begin{eqnarray}
\textmd{f}_p(z,a,b) \ = \ \frac{(a/b)^{p/2}}{2K_p(\sqrt{a b})} \ \! z^{p-1} \
\!\rme^{-(az + b/z)/2}\, ,
\label{23a}
\end{eqnarray}
is the generalized inverse Gaussian distribution~\cite{feller66}
($K_p$ is the modified Bessel function of the second kind with index
$p$). The LHS of (\ref{22a}) represents the PI for the free spinless
relativistic particle in the Newton--Wigner representation~\cite{NW}. The
full Klein--Gordon (KG) kernel which also contains the negative-energy spectrum can
be obtained from (\ref{22a}) by
considering the Feshbach--Villars representation of the KG equation and
making the substitution~\cite{JK2}
\begin{eqnarray}
\textmd{f}_{\frac{1}{2}}\!\left(\mathfrak{m}, tc^2, tc^2m^2\right) \!\mapsto \!\frac{1\! +\! \mbox{sgn}(t)\sigma_3}{2} \ \!  \textmd{f}_{\frac{1}{2}}\!\left(\mathfrak{m}, |t|c^2, |t|c^2m^2\right)\! .
\end{eqnarray}
The matrix nature of the smearing distribution ($\sigma_3$ is the Pauli matrix)
naturally includes the Feynman--Stuckelberg causal boundary condition and thus treats both particles and antiparticles in a symmetric way~\cite{JK2,Feshbach58}.
When the partition function is going to be calculated, the trace will get rid of the $\mbox{sgn}(t)$ term and $1/2$ is turned to $1$.

The explicit form of the identity (\ref{22a}) indicates that $\mathfrak{m}$ can be
interpreted as a Galilean-invariant Newton-like mass which takes on continuous values
distributed according to  $\textmd{f}_{\frac{1}{2}}\!\left(\mathfrak{m}, Tc^2, Tc^2m^2\right)$ 
with $\langle \tilde{m} \rangle = m + 1/Tc^2$ and
$\mbox{var}(\mathfrak{m}) = m/Tc^2 + 2/T^2c^4$.
Fluctuations of the Newtonian mass can be then depicted as originating
from particle's evolution in an inhomogeneous or granular  medium. Granularity, as known, for example, from
solid-state systems, typically leads to corrections in the local
dispersion relation~\cite{Johnson:93} and hence to alterations in
the local {\em effective mass}. The following picture thus emerges: on the
short-distance scale, a non-relativistic particle can be
envisaged as propagating via classical Brownian motion
through a single grain with a local mass $\mathfrak{m}$.
This fast-time process
has a time scale $\sim 1/{\mathfrak{m}}c^2$. An averaged
value of the local time scale
represents a transient temporal scale $\langle1/\mathfrak{m}c^2\rangle = 1/mc^2$
which coincides with particle's
Compton time $T_{_C}$ --- i.e., the
time for light to cross the particle's Compton wavelength.
At time scales much longer than $T_{_C}$ (large-distance scale),
the probability that the particle encounters
a grain which endows it with a mass ${\mathfrak{m}}$ is
$\textmd{f}_{\frac{1}{2}}\!\left(\mathfrak{m}, Tc^2, Tc^2m^2\right)$.
As a result one may
view a single-particle relativistic theory as a single-particle
non-relativistic theory where the particle's Newtonian mass ${\mathfrak{m}}$
represents a fluctuating parameter which approaches on average the
Einstein rest mass $m$ in the large $t$ limit.
We stress that $t$  should be understood
as the {\em observation time}, a time after which the observation
(position measurement) is made.
In particular, during the period $t$ the system remains unperturbed.
One can thus justly expect that in the long run all mass fluctuations will be
washed out and only a sharp time-independent effective mass will be perceived.
The form of $\langle {\mathfrak{m}} \rangle$ identifies
the time scale at which this happens with $ t \sim 1/mc^2$, i.e. with the Compton time $T_{_C}$.
It should be stressed that above mass fluctuations have nothing
to do with the {\em Zitterbewegung} which is caused by interference between positive- and negative-energy wave components. In our formulation both regimes are decoupled.

We may also observe that by coarse-graining the velocity over the
correlation time $T_{_C}$ we have
\begin{eqnarray}
\mbox{\hspace{-9mm}}&&\langle |{\bi v}| \rangle_{_{T_C}}   \ = \
\left.\frac{\langle |{\bi p}| \rangle}{\langle \mathfrak{m} \rangle} \right|_{_{T_C}} = \ c\, .
\end{eqnarray}
So on a short-time scale of order $\lambda_{_C}$ the spinless relativistic particle
propagates with an averaged velocity which is the speed of light $c$. But if one
checks the particle's position at widely separated intervals (much larger than $\lambda_{_C}$),
then many directional reversals along a typical PI trajectory will take place, and the particle's
net velocity will be then less than $c$ --- as it should be for a massive particle.
The particle then acquires a sharp
mass equal to Einstein's  mass, and the process (not being
hindered by fluctuating masses) is purely Brownian.
This conclusion is in line with the well-known Feynman
checkerboard picture~\cite{JS,Schulman-Jacobson} to which it reduces in the case of $(1+1)$D relativistic Dirac particle.

{\em {Robustness of emergent special relativity.}}~---~Understanding the robustness of the emergent Special Relativity under small variations
in the mass-smearing distribution function $\textmd{f}_{\frac{1}{2}}$ can guide the study of
the relation between Einsteinian SR and other deformed variants of SR, such as
Magueijo--Smolin and Amelino-Camelia's doubly special relativity~\cite{DSR,Kowalski}, or (quantum) $\kappa$-Poincar\'{e}
deformation of relativistic kinematics~\cite{Lukierski:08}.
In DRS models a further invariant scale $\ell$ is introduced, besides the usual speed of light $c$, and $\ell$ is typically considered to be of order of the Planck length. A small variation $\delta \textmd{f}_{\frac{1}{2}}$ of the smearing function originates the new Hamiltonian
\begin{eqnarray}
\bar{H} \ = \ \frac{\epsilon_1}{4} + \left(1 + \frac{\epsilon_0}{2}\right)\sqrt{{{\bi p}^2 c^2 + m^2 c^4  + \frac{\epsilon_2}{4}}}\; ,
\label{zaver2}
\end{eqnarray}
with $\epsilon_ 1  =  -2 \left(1 + \epsilon_0/2\right){\sqrt{\epsilon_2}}$ (see Ref.~\cite{JS} for details). By setting
\begin{eqnarray}
\epsilon_1 =  2 \left(\sqrt{\frac{1}{1-c^2 m^2 \ell^2}}\ \!  - \ \! 1 \right), \quad
\epsilon_2  =    \frac{4c^6 m^4 \ell^2}{1-c^2 m^2 \ell^2}\, , \nonumber
\end{eqnarray}
the new Hamiltonian $\bar{H}$ can be easily identified with
\begin{eqnarray}
\bar{H} \ = \ c\ \!\frac{{- m^2 c^2 \ell  \mp \sqrt{{\bi p}^2(1-m^2 c^2\ell^2) +
m^2 c^2}}}{1-m^2c^2\ell^2}\, ,
\label{HDSR1}
\end{eqnarray}
which coincides with the Magueijo--Smolin's doubly special relativistic Hamiltonian, in, say, its version~\cite{DSRII}.
It should be stressed that the Hamiltonian (\ref{zaver2}) (when also negative energy states are included) violates CPT symmetry. This is a typical byproduct of the Lorentz symmetry violation in many deformed SR systems.

For the Hamiltonian (\ref{HDSR1}) a relation analog to (\ref{22a}) holds, where now the smearing function has the form
$\textmd{f}_{\frac{1}{2}}\!\left(\mathfrak{m}, Tc^2\lambda, Tc^2m^2\lambda\right)$ with $\lambda=1/(1-m^2c^2\ell^2)$. The correlation distance
$1/mc\lambda$ can be naturally assumed as the minimal size $L_{_{\rm{GRAIN}}}$ of the ``grain of space" of the polycrystalline medium, which is linked to the new invariant scale $\ell$ by
\be
L_{_{\rm{GRAIN}}}:= \frac{1}{mc\lambda} = \lambda_C (1 - m^2c^2\ell^2)\,.
\ee
By tuning the size $L_{_{\rm{GRAIN}}}$ of these "grains of space" it is possible to pass continuously from Lorenz symmetry to other different symmetries, as those enjoyed by DSR models. We can in principle speculate that each large scale symmetry could originate from a specific kind of space(time) granularity.


{\em {Quantum field theory.}}~---~The superstatistics transition probability (\ref{6aa}) was constructed on the premise that ${H}$ is associated with a single particle. Of course, a single-particle relativistic quantum theory is logically untenable, since a multi-particle production is allowed whenever the particle reaches the threshold energy for pair production. In addition,
Leutwyler's no-interaction theorem~\cite{Leutwyler:65} prohibits interaction for any finite number of particles in the context of relativistic mechanics. To evade the no-interaction theorem it is necessary to have an infinite number of degrees of freedom to describe interaction. The latter is typically achieved via local quantum field theories (QFT).

It should be underlined in this context that the PI for a single
relativistic particle is still a perfectly legitimate building block even in QFT.
Recall that in the standard perturbative treatment of, say, {\em generating functional} for a {\em scalar field} each Feynman
diagram is composed of integrals over product of free correlation
functions (Feynman's correlators):
\begin{eqnarray}
\mbox{\hspace{-1.5mm}}\Delta_F({\bi y}, c t_y ; {\bi z }, c t_z) \! = \!  \frac{1}{4}\!\int_{-\infty}^{\infty} \!\!\!\!\!{\rmd\tau} \ \! \mbox{sgn}(\tau-t_y) \wp({\bi y}, \tau| {\bi z },t_z),
\label{zaver2b}
\end{eqnarray}
and may thus be considered as a functional of the PI $\wp({\bf x}',t'|{\bf x},t)$.
In fact, QFT in general, can be viewed as a grand-canonical ensemble of fluctuating particle histories (worldlines)
where Feynman diagrammatic representation of quantum fields depicts directly the pictures
of the world-lines in a grand-canonical ensemble.
This is the so-called ``worldline quantization" of particle
physics, and is epitomized, e.g., in Feynman's worldline representation of the one-loop affective action in quantum electrodynamics~\cite{Feynman:50}, in Kleinert's disorder field theory~\cite{KleinertIII}
or in the Bern--Kosower and Strassler ``string-inspired" approaches to QFT~\cite{Bern-Kosower}.

Because of (\ref{zaver2b}), the relationship between bosonic Bern--Kosower Green's function $G_B(\tau_1,\tau_2)$ and
the PI $\wp({\bf x}',t'|{\bf x},t)$ can be found easily through the known functional relation between $G_B$ and
$\Delta_F$, cf. Refs.~\cite{Bern-Kosower}.


{\em Gravity and Cosmology.} --- When spacetime is curved, a
metric tensor enters in both PI's in (7) in a different way,
yielding different ``counterterms"~\cite{PI,fiorenzo}. For instance,
in Bastianelli--van~Nieuwenhuizen's time slicing
regularization scheme~\cite{fiorenzo}
one has (when $\hbar$ is reintroduced)
\begin{eqnarray}
&&\mbox{\hspace{-8mm}}\frac{{\bi p}^2}{2
\mathfrak{m}} \ \mapsto \ \frac{{g^{ij}p_ip_j
}}{2
\mathfrak{m}}   + \frac{\hbar^2}{8\mathfrak{m}} (R + g^{ij}\Gamma_{il}^{m}\Gamma_{jm}^{l}
)\, , \nonumber \\[1mm]
&&\mbox{\hspace{-8mm}}\sqrt{{\bi p}^2 + m^2 c^2}  \mapsto  \sqrt{g^{ij}p_ip_j + \frac{\hbar^2}{4} (R + g^{ij}\Gamma_{il}^{m}\Gamma_{jm}^{l}
) + m^2 c^2}\nonumber \\
&&\mbox{\hspace{-8mm}} + \ \hbar^4 \Phi(R, \partial R, \partial^2R)
\ + \ \mathcal{O}(\hbar^6)\, ,\label{G.1.a}
\end{eqnarray}
where $g^{ij}$, $R$, $\Gamma_{kl}^{j}$  and $\Phi(\ldots)$ are the (space-like) pull-back metric tensor,
the scalar curvature, the Christoffel symbol, and non-vanishing function of $R$ and its first and second derivatives, respectively.
This causes the superstatistics identity
(7) to break down, as can be explicitly checked to the lowest order
in $\hbar$.
The respective two cases will thus lead
to different physics. Because the Einstein {\em equivalence principle} requires that the {\em local} spacetime
structure can be identified with the Minkowski spacetime possessing Lorentz symmetry, one might assume the
validity of (\ref{22a}) at least locally. However, in different space-time points
one has, in general, a different typical length scale of
the local inertial frames, depending on the gravitational field. 
The characteristic size of the local  inertial (i.e. Minkowski) frame is of order
$1/|K|^{1/4}$ where
$K = R_{\alpha\beta\gamma\delta}R^{\alpha\beta\gamma\delta}$ is the Kretschmann
invariant and $R_{\alpha\beta\gamma\delta}$ is the Riemann curvature.
Relation (\ref{22a}) tells us that the special relativistic description breaks down in regions of size smaller than $\lambda_{_C}$. For curvatures large enough, namely for strong gravitational fields, the size of the local inertial frame can become smaller than $\lambda_{_C}$, that is $1/|K|^{1/4}\lesssim \lambda_{_C}$. In such regions the special relativistic description is no more valid, and according to (\ref{22a}) must be replaced by a Newtonian description of the events.
For instance,
in Schwarzschild geometry we have $K=12\, r_s^2/r^6$, and the breakdown should
be expected at radial distances $r \lesssim (\lambda_{_C}^2
r_s)^{1/3}$ ($r_s$ is
the Schwarzschild radius) which are --- apart from
the hypothetical case of micro-black holes (where
$\lambda_{_C}\simeq r_s$) --- always deeply buried below the Schwarzschild event horizon.
In the cosmologically
relevant Friedmann-–Lema\^{\i}tre–-Robertson–-Walker  (FLRW) geometry, we have $K=12\,(\dot{a}^4 + a^2\ddot{a}^2)/(ac)^4$, and
the breakdown happens when $(\dot{a}^4 + a^2\ddot{a}^2) \gtrsim (ac/\lambda_{_C})^4$,  where $a(t)$ is the FLRW scale factor of the Universe and $\dot{a} = \rmd a/\rmd t$. Applying the well-known Vilenkin--Ford model~\cite{vilenkin} for inflationary cosmology, where $a(t)$ is given by: $a(t) = A\sqrt{\sinh(B t)}$ with $B = 2c\sqrt{\Lambda/3}$ ($\Lambda$ is the cosmological constant), we obtain a temporal bound on the validity of local Lorentz invariance, which, expressed in FLRW time, is
\begin{eqnarray}
t \ \lesssim \  \frac{1}{B} \ \!\mbox{arcsinh}\left[\frac{B\lambda_{_C}}{\left(8c^4-(B\lambda_{_C})^4\right)^{1/4}}
\right] \ \equiv \ \bar{t}\, .\label{G.2.a}
\end{eqnarray}
By using the presently known~\cite{lambda} value $\Lambda\simeq10^{-52}$m$^{-2}$ and the $\tau$-lepton Compton's wavelength $\lambda_{_C}^{\tau}\simeq6.7\times10^{-16}$m (yielding the tightest upper bound on $t$), we obtain $\bar{t} \simeq 4\times 10^{-24}$s.
Note that, since $B\lambda_{_C} \ll c$, then $\bar{t} \simeq \lambda_{_C}/c = t_{_C}$.
Such a violation of the local Lorentz invariance naturally breaks the particle-antiparticle symmetry since there is no unified theory of particles and antiparticles in the non-relativistic physics --- formally one has two separate theories. If the resulting matter-antimatter asymmetry provides a large enough CP asymmetry then this might have essential consequences in the early Universe, e.g., for leptogenesis. In this respect, $\bar{t}$ is
compatible with the {\em nonthermal} leptogenesis period that typically dates between $10^{-26}$--$10^{-12}$s after the Big Bang.


{\em {Conclusions and perspectives.}}~---~The new superstatistics PI representation
of a relativistic point particle introduced in this Letter, realizes an explicit
quantum mechanical duality between Einsteinian and Galilean relativity.
It also makes explicit how the SR invariance is encoded in the grain smearing distribution. Notably, the exact LS of a spacetime has no fundamental significance in our analysis, as it is only an accidental symmetry of the
coarse-grained configuration space in which a particle executes a standard Wiener process.
In passage from grain to grain particles's
Newtonian mass fluctuates according to an inverse Gaussian distribution.
The observed inertial mass of the particle is thus not a fundamental constant,
but it reflects the particle's interaction with the granular vacuum (cosmic field). This, in a sense, supports Mach's view of the phenomenon of inertia.

Interactions can be included in our framework in two different ways. The interaction with a background field
(such as electromagnetic field) can be directly treated with the superstatistics
prescription (\ref{22a}), see~\cite{JK2}. On the other hand, the multi-particle
interactions can be consistently formulated by ``embedding" the relativistic PI in QFT
via the worldline quantization. Such an embedding
may help to study several cosmological implications of systems
with granular space. If any of such systems quickly flows to the infrared fixed point, any direct effect due to the space discreteness, and related SR violation, might be insignificant on cosmological scales (where Lorentz and diffeomorphism invariance are restored),
while it might be crucial in the early Universe, e.g., for leptogenesis and the ensuing baryogenesis. Consequences on the detailed structure of the Cosmic Microwave Background spectrum will be explored in future work.

The presented approach implies a preferred frame. In this connection it is worth of noting
that, despite the fact that (\ref{22a}) is not manifestly LS invariant, one may
use the St\"{u}ckelberg trick and introduce a new fictitious variable
into the PI~(\ref{22a}), in such a way that the new action will have the
reparametrization symmetry, but will still be
dynamically equivalent to the original action. For relevant details see~Ref.~\cite{JS}. By not knowing the source,
one may then view this artificial gauge invariance as being a fundamental or even a defining property of the relativistic theory.
One might, however, equally well, proclaim the
``polycrystalline" picture as being a basic (or primitive) edifice of SR and view
the reparametrization symmetry as a mere artefact of an artificial redundancy that
is allowed in our description. It is this second view that we favored
here.

The presented scenario cannot directly accommodate the massless particles such as photons (identity (\ref{22a}) holds true only for $m\neq0$). One possibility would be to use the PI representation of Polyakov--Wheeler for massless particles and try to construct a similar superstatistics duality as in the case of massive particles.
This procedure is, however, not without technical difficulties and currently is under investigation. Conceptually is far more simpler to assume that the photon has a small mass. At present, there are a number of experimental limits to the mass of the photons~\cite{chibisov}. For instance, tests based on Coulomb's law and the galactic vector potential set the upper limit of $m_{\gamma} \lesssim 10^{-18}$ eV/c$^2$ $\simeq 10^{-57}$g. This gives the domain correlation distance for the photon $\simeq 1/m_{\gamma}c^2 \simeq 10^{43}$m  which is bigger than the radius of observable Universe ($\simeq 10^{26}$ m) and so in this picture the photon mass does not fluctuate --- it is a quasi-invariant.

Finally, this approach should reinforce the links between superstatistics paradigm
and the approach to quantum gravity based on stochastic quantization~\cite{Loll:09}.
In particular, the outlined granular space could be a natural model for the noise
terms in a Parisi--Wu stochastic-like quantization approach to gravity.

{\em {Acknowledgement.}}~---~
The writers are grateful to H.~Kleinert, Z.~Haba, M.~Sakellariadou,
and L.S.~Schulman for useful feedbacks.
This work was supported in part by GA\v{C}R Grant No. P402/12/J077.

\vspace{-6mm}


%
\end{document}